\documentstyle[twocolumn,aps,epsf]{revtex}

\newcommand{\operator}[1]{\hat{#1}}
\newcommand{\expect}[1]{\langle{#1}\rangle}
\newcommand{\boldvec}[1]{\mbox{\boldmath$#1$}}
\newcommand{\smallvec}[1]{\mbox{\boldmath$\scriptstyle#1$}}

\newcommand{\fieldop}[2]{\operator{\psi}_{#1}(\boldvec{#2})}
\newcommand{\fieldopdag}[2]{\operator{\psi}^{\dag}_{#1}(\boldvec{#2})}
\newcommand{\fluctop}[2]{\operator{\chi}_{#1}(\boldvec{#2})}
\newcommand{\fluctopdag}[2]{\operator{\chi}^{\dag}_{#1}(\boldvec{#2})}

\newcommand{\freeH}[2]{H_{#1}(\boldvec{#2})}
\newcommand{\meanH}[2]{{\bar\Upsilon}_{#1}(\boldvec{#2})}
\newcommand{\meanO}[1]{{\bar\Theta}(\boldvec{#1})}
\newcommand{\meanS}[1]{{\bar\Phi}(\boldvec{#1})}

\newcommand{\fluctHTot}[2]{{\tilde H}(\boldvec{#1},\boldvec{#2})}
\newcommand{\fluctH}[2]{{\tilde\Upsilon}(\boldvec{#1},\boldvec{#2})}
\newcommand{\fluctO}[2]{{\tilde\Theta}(\boldvec{#1},\boldvec{#2})}

\newcommand{\bigG}[2]{{\cal G}(\boldvec{#1},\boldvec{#2})}
\newcommand{\bigSigma}[2]{\Sigma(\boldvec{#1},\boldvec{#2})}

\newcommand{\kinetic}[2]{-{\hbar^2\over{#1}}\nabla^2_{\smallvec{#2}}}
\newcommand{\poten}[2]{V_{#1}(\boldvec{#2})}
\newcommand{\meanop}[2]{\phi_{#1}(\boldvec{#2})}
\newcommand{\meanstar}[2]{\phi^*_{#1}(\boldvec{#2})}
\newcommand{\GA}[2]{G_A(\boldvec{#1},\boldvec{#2})}
\newcommand{\GN}[2]{G_N(\boldvec{#1},\boldvec{#2})}

\newcommand{\UAA}{U_{\rm aa}}
\newcommand{\UAM}{U_{\rm am}}
\newcommand{\UMM}{U_{\rm mm}}

\begin{document}

\draft

\title{Formation of Pairing Fields in Resonantly Coupled Atomic and
  Molecular Bose-Einstein Condensates}

\author{M. Holland, J. Park, and R. Walser}

\address{JILA, University of Colorado and National Institute of
  Standards and Technology, Boulder, Colorado 80309-0440}

\date{\today}

\wideabs{

\maketitle

\begin{abstract}
  In this paper, we show that pair-correlations may play an important
  role in the quantum statistical properties of a Bose-Einstein
  condensed gas composed of an atomic field resonantly coupled with a
  corresponding field of molecular dimers. Specifically,
  pair-correlations in this system can dramatically modify the
  coherent and incoherent transfer between the atomic and molecular
  fields.
\end{abstract}


}

In quantum field theory, superfluidity and long-range order are
usually identified with a complex-valued order parameter defined as
the expectation value of the field operator. In the case of a dilute
Bose gas, the evolution of the order parameter is given in the
mean-field approximation by the well-known Gross-Pitaevskii equation.
Recently, this equation has been extensively applied to studies of the
zero-temperature collective behavior of metal-alkali Bose-Einstein
condensates confined in magnetic traps~\cite{stringari}.  Quantitative
agreement has been found between theory and experiment on a wide range
of distinct phenomena (e.g.\ Refs.~\cite{collective,vortex}).

Although identical in form to a non-linear Schr\"odinger equation, it
is notable that the Gross-Pitaevskii equation does not by itself
describe the full evolution of the quantum state. In general, a
complete description requires knowledge of the evolution of a
hierarchy of correlation functions, and not just the evolution of the
expectation value of the field. In particular, in low temperature
systems, it is well known that pairing fields associated with
two-particle correlations can play an important role. In a quantum
degenerate Fermi gas, pairing may radically alter the equilibrium
properties, giving rise to superfluidity in $^3$He and
superconductivity in electron systems. Correlations can also be
important in Bose systems. For example, squeezed states of
light---formed when photons are generated in pairs in nonlinear
media---have been studied extensively in the subject of quantum
optics~\cite{walls}.

\begin{figure}[b]
\begin{center}\
  \epsfysize=45mm \epsfbox{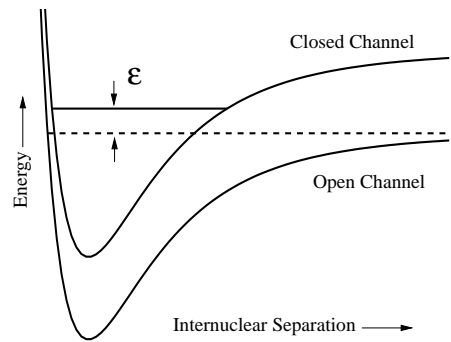}
\end{center}
\caption{
  Feshbach resonance. Atoms collide with relative kinetic energy near
  zero as indicated by the dotted line. This energy is
  quasi-degenerate with a bound state in a closed channel potential.
  Typically, $\epsilon$ can be tuned experimentally by changing the
  magnitude of an external magnetic field.}
\label{potens}
\end{figure}

In this paper, we investigate the role of pair-correlations on the
macroscopic dynamics of a dilute Bose-Einstein condensate composed of
atoms and molecules which are resonantly
coupled~\cite{tim,verhaar,nist,drummond}. Such a coupling may be
generated experimentally~\cite{mit,rb85} by tuning the strength of an
external magnetic field in the proximity of a Feshbach resonance (see
Fig.~\ref{potens}). Two parameters characterize the Feshbach
resonance: (i) the energy mismatch~$\epsilon$ between the bound state
and the zero energy edge of the continuum states for the colliding
atom pair, and (ii) the inverse lifetime~$\kappa$ of the bound state.
As an alternative, photoassociation may be used to directly generate
the resonant coupling~\cite{photo,heinzen}.  In photoassociation, a
two-photon Raman transition is used to couple the atomic continuum
states and a specific bound molecular level.  In that case $\epsilon$
represents the detuning energy of the Raman lasers from the
atom-molecule transition, and $\kappa$ denotes the two-photon Rabi
frequency proportional to the laser intensities and to the usual
overlap integrals.

An effective Hamiltonian for this coupled atom-molecule system may be
written as
\begin{eqnarray}
  \operator{H} &=& \int d^3\boldvec{x} \,
  \fieldopdag{a}{x}\freeH{a}{x}\fieldop{a}{x} +
  \fieldopdag{m}{x}\freeH{m}{x}\fieldop{m}{x}\nonumber\\
  &&{}+ {\UAA\over2} \int d^3\boldvec{x} \, \fieldopdag{a}{x}
  \fieldopdag{a}{x} \fieldop{a}{x}
  \fieldop{a}{x}\nonumber\\
  &&{}+ \UAM \int d^3\boldvec{x} \, \fieldopdag{a}{x}
  \fieldopdag{m}{x} \fieldop{m}{x}
  \fieldop{a}{x}\nonumber\\
  &&{}+ {\UMM\over2} \int d^3\boldvec{x} \, \fieldopdag{m}{x}
  \fieldopdag{m}{x} \fieldop{m}{x}
  \fieldop{m}{x}\nonumber\\
  &&{}+ {g\over2} \int d^3\boldvec{x} \, \fieldopdag{m}{x}
  \fieldop{a}{x} \fieldop{a}{x}+\mbox{h.c.}\,,
\label{effective}
\end{eqnarray}
where $\fieldop{a}{x}$ and $\fieldop{m}{x}$ are bosonic field
operators which annihilate an atom or molecule respectively at
coordinate~$\boldvec{x}$. Atomic binary collisions are characterized
by the mean-field energy per unit density, $\UAA=4\pi\hbar^2a/m$,
where $m$ is the atomic mass, and $a$ is the scattering length for
atom-atom collisions. Equivalent definitions apply for $\UAM$ and
$\UMM$ corresponding to atom-molecule collisions and molecule-molecule
collisions. The process of inter-conversion of atom pairs into
molecules is characterized by $g=\sqrt{\kappa\,\UAA}$~\cite{tim}.  The
free Hamiltonians are
\begin{eqnarray}
  \freeH{a}{x}&=&\kinetic{2m}{x} + \poten{a}{x} - \mu_a\nonumber\\
  \freeH{m}{x}&=&\kinetic{4m}{x} + \poten{m}{x} - \mu_m\,,
\end{eqnarray}
where we include $\poten{a}{x}$ and $\poten{m}{x}$ for generality to
allow for the possibility of external potentials. The chemical
potentials of the atomic and molecular fields are denoted by $\mu_a$
and $\mu_m$ respectively, such that $\epsilon=\mu_m-2\mu_a$.

We solve for the evolution of the fields, assuming the principle of
attenuation of many-particle correlations in the dilute gas. That is,
we find the expectation value of the Heisenberg equations for
$\fieldop{a}{x}$ and $\fieldop{m}{x}$ and separate explicitly the mean
values of the field operators
\begin{eqnarray}
  \meanop{a}{x} &=& \expect{\fieldop{a}{x}} \nonumber\\
  \meanop{m}{x} &=& \expect{\fieldop{m}{x}}
\end{eqnarray}
from their zero-mean fluctuating components
\begin{eqnarray}
  \fluctop{a}{x} &=& \fieldop{a}{x}-\meanop{a}{x} \nonumber\\
  \fluctop{m}{x} &=& \fieldop{m}{x}-\meanop{m}{x}.
\end{eqnarray}
In order to close the dynamic equations, we expand products of three
of more fluctuating operators using Wick's theorem (i.e.\ we assume a
Gaussian reference distribution for the many-body quantum state).
Since we intend to determine the pairing statistics of atoms and do
not anticipate intrinsic pairing of molecules to play a significant
role, we assume a classical field for the molecular quantum state
(i.e.\ we specify the molecular field as an exact eigenstate of
$\fieldop{m}{x}$). Following this procedure, we derive a description
of the coupled atom-molecule system, which involves the following
dynamical quantities: the condensates $\meanop{a}{x}$ and
$\meanop{m}{x}$, and the atomic fluctuations as described by normal
densities $\GN{x}{y}$, and anomalous densities $\GA{x}{y}$, defined by
\begin{eqnarray}
  \GN{x}{y} &=& \expect{\fluctopdag{a}{y}\fluctop{a}{x}}
  \nonumber\\
  \GA{x}{y} &=& \expect{\fluctop{a}{y}\fluctop{a}{x}}.
\end{eqnarray}
Following this prescription, we arrive at the following dynamical
equations for the condensates:
\begin{eqnarray}
  i\hbar{d\meanop{a}{x}\over dt} &=&
  \bigl(\freeH{a}{x}+\meanH{a}{x}\bigr)\meanop{a}{x} +
  \meanO{x}\meanstar{a}{x} \nonumber\\
  i\hbar{d\meanop{m}{x}\over dt} &=&
  \bigl(\freeH{m}{x}+\meanH{m}{x}\bigr)\meanop{m}{x} + \meanS{x}\, ,
\label{meanfldeq}
\end{eqnarray}
with corresponding mean-field potentials
\begin{eqnarray}
  && \meanH{a}{x} = \UAA\bigl(|\meanop{a}{x}|^2 + 2\GN{x}{x}\bigr) +
  \UAM|\meanop{m}{x}|^2 \nonumber\\
  && \meanH{m}{x} = \UAM\bigl(|\meanop{a}{x}|^2 + \GN{x}{x}\bigr) +
  \UMM|\meanop{m}{x}|^2\, ,
\end{eqnarray}
and coupling elements
\begin{eqnarray}
  \meanO{x} &=& \UAA\GA{x}{x} + g\meanop{m}{x} \nonumber\\
  \meanS{x} &=& {g\over2}\bigl(\meanop{a}{x}^2+\GA{x}{x}\bigr)\,.
\end{eqnarray}
A concise representation for the evolution of the atomic fluctuations
is given by
\begin{equation}
  i\hbar{d{\cal G}\over dt} = \Sigma{\cal G} - {\cal
    G}\Sigma^{\dag}\,,
\label{densitymatrixeq}
\end{equation}
which is written in terms of the matrix
\begin{equation}
  \bigG{x}{y}=\left(\begin{array}{cc}
      \expect{\fluctopdag{a}{y}\fluctop{a}{x}} &
      \expect{\fluctop{a}{y}\fluctop{a}{x}} \\
      \expect{\fluctopdag{a}{y}\fluctopdag{a}{x}} &
      \expect{\fluctop{a}{y}\fluctopdag{a}{x}}
\end{array}\right)\,,
\end{equation}
and Bogoliubov self-energy of general structure
\begin{eqnarray}
\bigSigma{x}{y}&=&\left(
\begin{array}{cc}
  \fluctHTot{x}{y} & \fluctO{x}{y} \\
  -\fluctO{x}{y}^* & -\fluctHTot{x}{y}^*
\end{array}
\right)\,.
\label{fluct}
\end{eqnarray}
The $2\times2$ elements of $\bigSigma{x}{y}$ are each diagonal in
$\boldvec{x}$ and $\boldvec{y}$ due to the assumption of a contact
potential for all scattering diagrams in Eq.~(\ref{effective}).
Accordingly, by defining
$\fluctHTot{x}{x}=\freeH{a}{x}+\fluctH{x}{x}$, the full solution is
\begin{eqnarray}
  \fluctH{x}{x}&=& 2\UAA\bigl(|\meanop{a}{x}|^2 + \GN{x}{x}\bigr) +
  \UAM|\meanop{m}{x}|^2\nonumber\\
  \fluctO{x}{x}&=&\UAA\bigl(\meanop{a}{x}^2+\GA{x}{x}\bigr)
  +g\meanop{m}{x}.
\end{eqnarray}

This forms a universal description of the Hartree-Fock-Bogoliubov
theory of the coupled atom-molecule condensate system. As an example
of the implementation of these equations, we now apply the theory to
the case of a uniform gas with the assumption that we may ignore the
mean-field potentials (i.e.\ we set to zero $\UAA$, $\UAM$, and
$\UMM$), along with the external potentials $\poten{a}{x}$ and
$\poten{m}{x}$. Due to translational symmetry, the resulting atomic
and molecular condensates are spatially homogeneous, and the normal
and anomalous densities, $\GN{x}{y}$ and $\GA{x}{y}$, depend only on
the magnitude of the relative coordinate, i.e.\
$|\boldvec{x}-\boldvec{y}|$.

We convert the equations to dimensionless form in the following
manner. We define the number density $n$ as the number of atoms plus
twice the number of molecules per unit volume. Then $g\sqrt{n}$
denotes a characteristic coupling energy of the atomic and molecular
mean-fields. A dimensionless time $\tau$ can therefore be defined by
$\tau=g\sqrt{n}\,t/\hbar$. A dimensionless coordinate
$r=|\boldvec{x}-\boldvec{y}|/2\zeta$ is associated with the length
scale $\zeta$ corresponding to the formation of molecules from atom
pairs. This is effectively a ``healing length'' for molecule formation
found by balancing the relative kinetic energy with the resonance
coupling energy, i.e.\ $\hbar^2/2\mu\zeta^2=g\sqrt{n}$ where $\mu=m/2$
is the reduced mass.  Making systematic substitutions
$\phi_a=\meanop{a}{x}/\sqrt{n}$, $\phi_m=\meanop{m}{x}/\sqrt{n}$,
$G_N(r)=\GN{x}{y}/n$, $G_A(r)=\GA{x}{y}/n$, and
$\Delta=\epsilon/g\sqrt{n}$ gives the complete system of equations in
dimensionless form
\begin{eqnarray}
  i{d\phi_a\over d\tau}&=&\phi_a^*\phi_m\nonumber\\
  i{d\phi_m\over d\tau}&=&-\phi_m\Delta
  +{1\over2}\left(\phi_a^2+G_A(r)|_{r=0}\right)
  \nonumber\\
  i{dG_N(r)\over d\tau}&=&\phi_mG_A^*(r)-\phi^*_mG_A(r)\nonumber\\
  i{dG_A(r)\over d\tau}&=&-\nabla_r^2G_A(r)
  +\phi_m\Bigl(2G_N(r)+\eta\,\delta^{(3)}(r)\Bigr)\,,
\label{uniform}
\end{eqnarray}
where $\delta^{(3)}(r)$ is the isotropic three-dimensional Dirac delta
function and we have defined $\eta=1/n\zeta^3$ as the inverse
diluteness parameter of the coupled atomic and molecular gas. There
are explicitly conserved quantities in these equations corresponding
to normalization (density of atoms plus twice the density of
molecules) i.e.\
\begin{equation}
|\phi_a|^2+G_N(r)|_{r=0}+2|\phi_m|^2=1\,,
\end{equation}
and energy density $u$ (energy due to detuning and coupling plus the
kinetic energy of the normal gas)
\begin{eqnarray}
  u&=&-\Delta|\phi_m|^2+{\rm Re}
  \left[\phi_m^*(\phi_a^2+G_A(r)|_{r=0})\right]\nonumber\\
  &&{}-{1\over2}\left.\nabla_r^2G_N(r)\right|_{r=0}\,.
\end{eqnarray}

Interestingly, Eqns.~(\ref{uniform}) cannot be directly integrated as
written. The delta function describes the {\em spontaneous} breakup of
molecules into atom pairs.  This process is reminiscent of the decay
of an excited atom leading to the spontaneous emission of a photon. In
that case, the interaction with the unoccupied vacuum modes of the
radiation field gives both an energy width and an energy shift (Lamb
shift) to the decaying state.  Analogously in the atom-molecule
system, there is an energy shift of the molecular state containing
contributions from all diagrams representing the breakup of molecules
into {\em virtual} atom pairs.  The virtual atom pairs may form an
intermediate state off the energy shell. Summing over the full
spectrum of wave numbers for the pair gives an infinite shift to the
molecular level.

This divergence is reconciled by noting that we have been inconsistent
in not including the self-energy of the molecular quantum state in the
original definition of the chemical potential $\mu_m$. We carry out
the renormalization in the following manner. We place an artificial
bound on the momenta of the atom pair by replacing the delta function
by a three dimensional Gaussian of standard width $\sigma_r$
normalized to have unit volume
\begin{equation}
  \delta^{(3)}(r)\rightarrow(2\pi\sigma_r^2)^{-{3\over2}}
  e^{-r^2/2\sigma_r^2}\,.
\label{deltasub}
\end{equation}
Physically, this accounts for the fact that the real potential is not
exactly a contact potential. The resulting energy shift is then finite
and we derive its value according to the following prescription. We
define a three dimensional Fourier transform
\begin{equation}
  G_A(k)=\int
  d^3\boldvec{r}\,G_A(r)e^{-i\smallvec{r}\cdot\smallvec{k}} \,,
\end{equation}
which for the isotropic case is simplified to
\begin{equation}
  G_A(k)={4\pi\over k}\int_0^{\infty}r\,G_A(r)\sin(kr)\,dr\,.
\end{equation}
For sufficiently high wave numbers, an approximate equation may then
be written for $G_A(k)$ using Eq.~(\ref{uniform})
\begin{equation}
  i{dG_A(k)\over d\tau}=k^2G_A(k)+\eta(k)\phi_m\,,
\end{equation}
where $\eta(k)=\eta\exp(-\sigma_r^2k^2/2)$. Taking $G_A(k)|_{\tau=0}=0$,
the solution of this equation is
\begin{eqnarray}
  G_A(k)&=&-i\eta(k)\int_0^{\tau}\phi_m(\tau')e^{ik^2(\tau'-\tau)}
  d\tau'\nonumber\\
  &\approx&-{1\over k^2}\eta(k)\phi_m(\tau)\,.
\end{eqnarray}
Calculating the energy shift of the molecular level requires
substituting this result into the evolution equation for
$d\phi_m/d\tau$ in Eq.~(\ref{uniform}). Using the Fourier integral
$G_A(r)|_{r=0}=(2\pi^2)^{-1}\int_0^{\infty}k^2G_A(k)\,dk$, the
resulting shift may be incorporated into a renormalized detuning
$\Delta$
\begin{equation}
  \Delta\rightarrow\Delta-{\eta\sqrt{2\pi}\over8\pi^2\sigma_r}\,.
\label{musub}
\end{equation}
The end result of this analysis is that Eqns.~(\ref{uniform}) are
satisfactorily renormalized by making the two substitutions,
Eq.~(\ref{deltasub}) and Eq.~(\ref{musub}).  The evolution of any
relevant observable will then be independent of the choice of
$\sigma_r$, providing $\sigma_r$ is chosen to be sufficiently small.

\begin{figure}
\begin{center}\
  \epsfysize=70mm \epsfbox{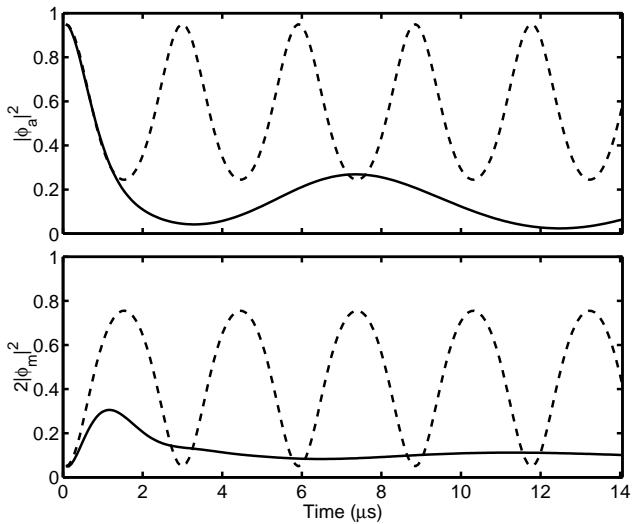}
\end{center}
\caption{
  Time evolution of the atomic condensate $|\phi_a|^2$ and molecular
  condensate $2|\phi_m|^2$ for solely mean-field theory (dashed) and
  for the Hartree-Fock-Bogoliubov theory (solid).}
\label{condensate}
\end{figure}

As a numerical example, we examine the $^{23}$Na Feshbach resonance at
907~G~\cite{mit}. For this system $\kappa=2\pi\hbar\times44$~MHz, and
we consider the bound state to be on-resonance with the colliding atom
pairs. Other parameters are the scattering length $a=60\,a_0$, where
$a_0$ is the Bohr radius, and we take a density of $10^{15}\,{\rm
  cm}^{-3}$. This gives an inverse diluteness parameter of
$\eta=16.5$. In the simulation in Fig.~\ref{condensate} we show the
evolution of the atomic and molecular condensate fields, comparing the
Hartree-Fock-Bogoliubov theory we have derived with the predictions of
mean-field theory. The initial condition is taken as 95\% of the
population in the atomic condensate and 5\% in the molecular
condensate. The persistent large scale oscillations of the population
of the atomic and molecular condensates as seen in a solely mean-field
theory dampen out when the coupling to the normal gas is included.
This is partly due to the fact that when molecules are formed from
atom pairs, the pair correlation function at that point is depleted,
and partly due to the possibility for spontaneous breakup of the
generated molecules into the normal component of the gas.

\begin{figure}
\begin{center}\
  \epsfysize=60mm \epsfbox{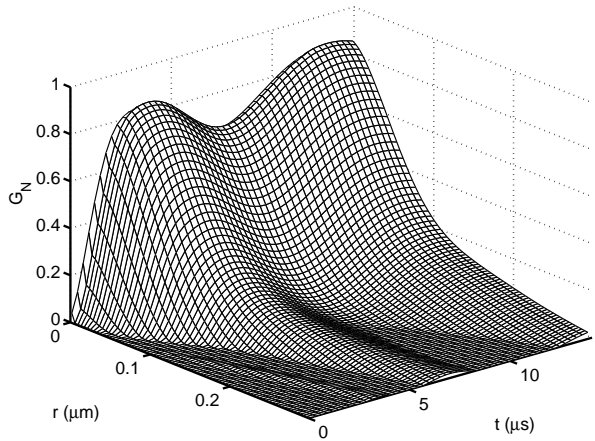}
\end{center}
\caption{Evolution of the normal density.}
\label{gn}
\end{figure}

In Fig.~\ref{gn}, we illustrate the behavior of the normal component
during this simulation. The density of the normal gas is
$G_N(r)|_{r=0}$. The temperature of the normal gas is found from the
spatial scale over which this correlation function decays. For
reference, for an equilibrium classical gas, $G_N$ is Gaussian with a
standard deviation given by the thermal de~Broglie wavelength.

\begin{figure}
\begin{center}\
  \epsfysize=60mm \epsfbox{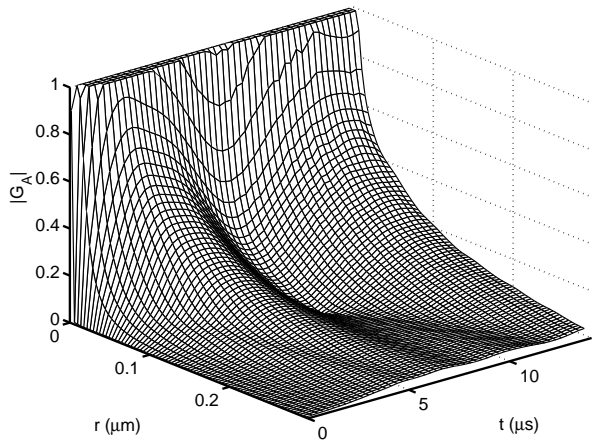}
\end{center}
\caption{Evolution of the anomalous density.}
\label{ga}
\end{figure}

In Fig.~\ref{ga}, we show the evolution of the anomalous fluctuations
for this example. For a classical gas, the anomalous fluctuations are
explicitly zero, so clearly the quantum statistics of the normal
component generated here are not the usual classical thermodynamic
equilibrium. One of the reasons for this is that there can never be an
odd number of atoms in the thermal cloud since atoms are spontaneously
generated from molecules in pairs. This situation is similar to the
formation of a squeezed vacuum in optics using a laser pump to drive a
parametric amplifier in order to produce photon pairs. Note the value
of $|G_A(r)|$ near $r=0$ depends on $\sigma_r^{-1}$ and is not an
observable.

In conclusion, the damping of the coherent atom-molecule oscillations
is accompanied by an increase in density of the normal component. The
behavior shown in Fig.~\ref{condensate} resolves the conceptual
difficulty associated with the rapid atom-molecule oscillations
predicted by mean-field theory---how can pairs of atoms find each
other rapidly enough to form molecules at the predicted rate? The
proper treatment of correlations shows that the oscillations are
slower in frequency and moreover rapidly damped.

We would like to thank J. Cooper, E. Cornell, C. Wieman, and A.
Andreev for helpful discussions. This work was supported by the
Department of Energy.


\begin{references}
\vspace*{-2pc}
  
\bibitem{stringari} F. Dalfovo, S. Giorgini, L. P. Pitaevskii, and S.
  Stringari, Rev.\ Mod.\ Phys.\ {\bf 71}, 463 (1999).
  
\bibitem{collective} M. Edwards, P. A.  Ruprecht, K. Burnett, R. J.
  Dodd, and C. W. Clark, Phys.\ Rev.\ Lett.\ {\bf 77}, 1671 (1996).
  
\bibitem{vortex} J. E. Williams and M. J. Holland, Nature {\bf 401},
  568 (1999); M. R. Matthews {\em et al.}, Phys.\ Rev.\ Lett.\ {\bf
    83}, 2498 (1999).
  
\bibitem{walls} D. F. Walls, Nature {\bf 324}, 210 (1986).
  
\bibitem{tim} E. Timmermans, P. Tommasini, R. C\^ot\'e, M. Hussein,
  and A. Kerman, Phys.\ Rev.\ Lett.\ {\bf 83}, 2691 (1999); E.
  Timmermans {\em et al.}, Phys.\ Rep.\ {\bf 315} 199 (1999);

\bibitem{verhaar} F. A. Abeelen and B. J. Verhaar, Phys.\ Rev.\ Lett.,
  {\bf 83} 1550 (1999).
  
\bibitem{nist} V. A. Yurovsky, A. Ben-Reuven, P. S. Julienne, and C.
  J. Williams, cond-mat/9903414.
  
\bibitem{drummond} P. D. Drummond, K. V. Kheruntsyan, H. He, Phys.\ 
  Rev.\ Lett.\ {\bf 81}, 3055 (1998).

\bibitem{mit} S. Inouye {\em et al.}, Nature {\bf 392}, 151 (1998); J.
  Stenger {\em et al.}, Phys.\ Rev.\ Lett. {\bf 82}, 2422 (1999).
  
\bibitem{rb85} Ph. Courteille {\em et al.}. Phys.\ Rev.\ Lett.\ {\bf
    81}, 69 (1998); J. L. Roberts {\em et al.}, {\em ibid.} {\bf 81},
  5109 (1998).
  
\bibitem{photo} P. S. Julienne {\em et al.}, Phys.\ Rev.\ A {\bf 58},
  R797 (1998).

\bibitem{heinzen} R. Wynar, R. S. Freeland, D. J. Han, C. Ryu, D. J.
  Heinzen {\em et al.},  Science {\bf 287}, 1016 (2000).

\end{references}
\end{document}